\documentclass[%
reprint,
superscriptaddress,
amsmath,amssymb
aps,
twocolumn,epsf,prl,
]{revtex4-2}

\usepackage{graphicx}% Include figure files
\usepackage{dcolumn}% Align table columns on decimal point
\usepackage{bm}% bold math
\usepackage{amsmath}
\usepackage{amsfonts}
\usepackage{amssymb}
\usepackage[dvipsnames]{xcolor}
\usepackage{hyperref}
\usepackage{pbox}
\usepackage{pinlabel}
\usepackage[justification=RaggedRight]{caption}
\usepackage[singlelinecheck=off]{subcaption}
\usepackage{ulem}
\usepackage{xr}
\begin{document}

	\title{Mott insulators with boundary zeros}
	
	\author{N.~Wagner}
	\affiliation{Institut f\"ur Theoretische Physik und Astrophysik, Universit\"at W\"urzburg, 97074 W\"urzburg, Germany}
	\author{L.~Crippa}%
	\affiliation{Institut f\"ur Theoretische Physik und Astrophysik and W\"urzburg-Dresden Cluster of Excellence ct.qmat, Universit\"at W\"urzburg, 97074 W\"urzburg, Germany}
	\author{A.~Amaricci}
	\affiliation{CNR-IOM, Istituto Officina dei Materiali, Consiglio Nazionale delle Ricerche, Via Bonomea 265, 34136 Trieste, Italy}
        \author{P.~Hansmann}
	\affiliation{Department of Physics, Friedrich-Alexander-Universit\"at Erlangen-N\"urnberg, 91058, Erlangen, Germany}
	\author{M.~Klett}
        \affiliation{Max-Planck-Institut f\"ur Festk\"orperforschung, Heisenbergstr. 1, 70569 Stuttgart, Germany}
	\author{E.~J.~K\"onig}
        \affiliation{Max-Planck-Institut f\"ur Festk\"orperforschung, Heisenbergstr. 1, 70569 Stuttgart, Germany}
	\author{T.~Sch\"afer}
        \affiliation{Max-Planck-Institut f\"ur Festk\"orperforschung, Heisenbergstr. 1, 70569 Stuttgart, Germany}
	\author{D.~Di~Sante}
	\affiliation{Department of Physics and Astronomy, University of Bologna, Bologna, Italy and CCQ-Flatiron Institute, New York, NY, USA}
        \author{J.~Cano}
        \affiliation{Department of Physics and Astronomy, Stony Brook University, Stony Brook, New York 11974, USA and CCQ-Flatiron Institute, New York, NY, USA}
	\author{A.~J.~Millis}
	\affiliation{Department of Physic, Columbia University, New York, NY, USA and Center for
Computational Quantum Physics, Flatiron Institute, New York, NY, USA}
	\author{A.~Georges}
\affiliation{Coll\`{e}ge de France, PSL University, 11 place Marcelin Berthelot, 75005 Paris, France}
\affiliation{Center for Computational Quantum Physics, Flatiron Institute, New York, New York 10010, USA}
\affiliation{Department of Quantum Matter Physics, University of Geneva, 24 quai Ernest-Ansermet, 1211 Geneva, Switzerland}
\affiliation{CPHT, CNRS, \'{E}cole Polytechnique, IP Paris, F-91128 Palaiseau, France}
	\author{G.~Sangiovanni}
  \email{sangiovanni@physik.uni-wuerzburg.de}
	\affiliation{Institut f\"ur Theoretische Physik und Astrophysik and W\"urzburg-Dresden Cluster of Excellence ct.qmat, Universit\"at W\"urzburg, 97074 W\"urzburg, Germany}

\maketitle

{\bf The topological classification of electronic band structures is based on symmetry properties of Bloch eigenstates of single-particle Hamiltonians.
In parallel, topological field theory has opened the doors to the formulation and characterization of non-trivial phases of matter driven by strong electron-electron interaction.
Even though important examples of topological Mott insulators have been constructed, the relevance of the underlying non-interacting band topology to the physics of the Mott phase has remained unexplored. 
Here, we show that the momentum structure of the Green's function zeros defining the ``Luttinger surface" provides a  topological characterization of the Mott phase  related, in the simplest description, to the one of the single-particle electronic dispersion.
Considerations on the zeros lead to the prediction of new phenomena: a topological Mott insulator with an inverted gap for the bulk zeros must possess gapless zeros at the boundary, which behave as a form of ``topological antimatter'' annihilating conventional edge states.
Placing band and Mott topological insulators in contact produces distinctive observable signatures 
at the interface, revealing the otherwise spectroscopically elusive Green's function zeros.}
\begin{figure*}[th]
\begin{center}
	\includegraphics[width=0.99\linewidth]{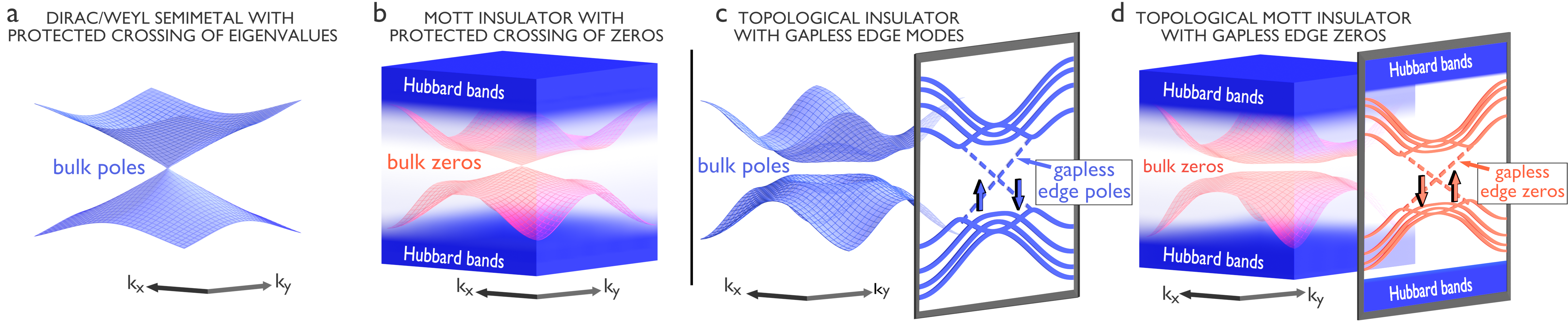}
\caption{{\bf Bulk and edge zeros in strongly correlated topological phases.} {\bf a}-{\bf b}, MIs originating at large values of $U$ from a semimetal with protected crossings and, {\bf c}-{\bf d}, from a conventional non-interacting TI.
In both situations the non-interacting dispersion represents the energy-momentum location of the Green's function poles (light blue surfaces). In the corresponding Mott phases, we find dispersive bulk zeros (red surfaces) inside the gap between the lower and upper Hubbard bands (dark blue). In the case of the semimetallic bandstructure, {\bf b}, the bulk zeros display a symmetry-protected crossing. For the topological insulating bandstructure, {\bf d}, the bulk zeros form instead an inverted gap, inherited from the non-interacting topological band structure. Therefore, gapless edge zeros appear, as indicated by the red dashed lines. Note the opposite spin direction associated to the edge zeros of the TMI w.r.t. to that of the edge states of the TI.}
	\label{fig:sketch}
\end{center}
\end{figure*}

\textbf{\large Introduction}\\
The theoretical description of topological order in physical systems has progressed along the parallel routes of band- and quantum field-theory \cite{fu_time_2006,schnyder_classification_2008,qi_topological_2008,wang_topological_2010,qi_topological_2011,bradlyn_topological_2017,kruthoff_topological_2017}.
The former, based on single-particle Hamiltonians, offers a clear explanation of the origin of topological invariants but is limited to the realm of weakly-interacting perturbation theory. 
The latter, making use of the Green's function formalism, encompasses a wider range of phases, such as topological Mott insulators \cite{raghu_topological_2008,wen_interaction-driven_2010,pesin_mott_2010}, at the cost of a higher theoretical complexity and a greater computational effort. 
This more sophisticated approach is, however, necessary since single-particle wave functions are no longer eigenvectors of the interacting many-electron Hamiltonian. 

Mott insulators (MIs) are characterized by an interaction-driven gap opening occurring without explicit breaking of an underlying symmetry or long-range ordering. 
Phenomena of this kind, which are intrisically non-perturbative in the coupling constant \cite{gunnarsson_breakdown_2017}, result in gaps even when globally robust crossings of bands such as Weyl cones or topological boundary states  \cite{bradlyn_beyond_2016,muechler_topological_2016,young_dirac_2012,kim_dirac_2015,wieder_double_2016} are present. 
In these cases \cite{wagner_resistivity_2021,braguta_monte_2016} as well as in MIs arising from conventional topological insulators (TIs) with an inverted gap \cite{budich_fluctuation-induced_2012,amaricci_first-order_2015}, a key question is if and how the topological information encoded in the non-interacting electronic dispersion survives and reemerges after the Mott transition.
Topological gapless excitations can occur in the context of MI quantum spin liquids 
concomitant with non-trivial entanglement of the fractionalized ground state \cite{savary_quantum_2017} allowing to circumvent the necessity of a Luttinger's theorem-imposed Fermi sea volume \cite{oshikawa_topological_2000,paramekanti_extending_2004}. 
Strongly correlated counterparts to bulk semimetals
disclose a relation to lattice symmetries in the case of non-symmorphic space groups 
\cite{watanabe_filling_2015,kimchi_featureless_2013,ware_topological_2015}.
Topological Luttinger invariants have been formulated specifically for such systems 
\cite{parameswaran_topological_2019}, broadening the conceptual perimeter of the Luttinger surface in a MI  \cite{dzyaloshinskii_consequences_2003}.
Further, a discrete symmetry-breaking has been proposed to be associated to a %the 
Mott-like transition \cite{huang_discrete_2022} considering an interaction term diagonal in momentum space \cite{morimoto_weyl_2016}. Recently, the existence of quasiparticles approaching a
Luttinger surface has been demonstrated \cite{fabrizio_emergent_2022,blason_unified_2023}.

Here, we search for symmetry-protected gapless modes despite the presence of a hard Mott gap and for experimentally observable fingerprints thereof.
We present analytic as well as numerical evidence that the zeros of the single-particle Green's functions have a dispersion in momentum space that can be topologically classified  as the corresponding non-interacting single-particle band structure.
Schemes based on the two-particle level have also been proposed \cite{he_topological_2016,soldini_interacting_2023,herzog-arbeitman_interacting_2022}.
Our approach offers a clean and easily accessible signature of the topology  
 of Mott insulating states: first of all, the bulk Green's function zeros of MIs arising from symmetry-protected semimetals  form Dirac and Weyl points and are thereby topologically distinct from ordinary ones.
Second, the zeros of a MI that originates  from a gapped bandstructure are in general also gapped. When such  bulk gap is of inverted character, exotic gapless zeros localized at the boundaries must exist.
We show their existence in various systems
 and propose how to detect them in an experiment, relying on the intrinsically incoherent state that forms when a pole and a boundary zero annihilate. 

\bigskip
\noindent	
\textbf{\large Results and Discussion}\\
{\bf Momentum dispersion of Green's function zeros} -- A pole of the single-particle propagator close to the real-frequency axis describes a conventional quasiparticle excitation of the system. 
A zero eigenvalue of $G({\bf k},\omega)$, with ${\bf k}$ and {$\omega$} indicating crystal momentum and complex frequency respectively, corresponds instead to a divergence of the self-energy $\Sigma({\bf k},\omega)$ and causes the opening of a Mott gap \cite{stanescu_fermi_2006,stanescu_theory_2007,sakai_evolution_2009,sakai_doped_2010}.
For an isolated atom, the pole of $\Sigma$ is momentum-independent and the corresponding continued fraction representation contains one floor only. Moving away from this extreme limit by switching on a finite hopping $t$, information on the lattice dispersion enters into the picture.
Yet, if we are deep in the Mott phase, the self-energy can be  expressed as :
\begin{align}
    \Sigma({\bf k},\omega)  =  \frac{U^2/4}{ \omega + \widetilde{H_0}({\bf k})},
    \label{eq:Sigma}
\end{align}
where $\widetilde{H_0}({\bf k})$ indicates the non-interacting  Hamiltonian with renormalized parameters. Eq.~\ref{eq:Sigma} is the result of a $t/U$ expansion in the presence of a Mott gap, which makes the lower floors of the continued fraction contribute with terms not larger than $t^2/U^2$. In particular, these terms can contain frequency-dependent real parts whereas the imaginary part of the self-energy has to vanish inside the Mott gap except at isolated poles. According to Eq. \eqref{eq:Sigma} the position of these poles at large $U/t$ is given by $-\widetilde{H_0}({\bf k})$. In the Supplementary Information \cite{noauthor_supplementary_2023} we describe how the renormalization encoded in $\widetilde{H_0}({\bf k})$ depends on the spatial locality of the various terms of the tight-binding model and how it can be obtained in a controlled way through a calculation of  the  spin-density correlators on different sites $\langle  n_{i\sigma}n_{j\sigma^\prime} \rangle$
 Their values are beyond mean field because of the \textbf{k}-dependence of our self-energy \cite{rohringer_impact_2016}. 
 
Expressions similar to Eq.~\ref{eq:Sigma} have been discussed previously \cite{nolting_methode_1972,harris_single-particle_1967,roth_electron_1969,onoda_mott_2003,avella_composite_2012,avella_hubbard_2014}, for one-dimensional systems \cite{berthod_breakup_2006}, in single-orbital models \cite{pairault_strong-coupling_2000}, in the pseudogap phase \cite{pudleiner_momentum_2016}, to prove the breakdown of Luttinger's theorem in a MI \cite{rosch_breakdown_2007} as well as in the context of magnetically ordered phases \cite{altshuler_luttinger_1998} and doped spin liquids \cite{konik_doped_2006}.

In this respect, it is interesting to mention also the so-called ``failed'' (quantum disordered) 
superconductors or spin density-wave systems, in which fermions couple to a strongly fluctuating boson which prevents long-range ordering. In the latter case, the $+$ sign is a consequence of $H_0({\bf k}+{\bf Q})=-H_0({\bf k})$ appearing in the denominator \cite{altshuler_luttinger_1998}. This observation suggests an intimate connection between Eq.~\ref{eq:Sigma} and paradigmatic quantum spin liquids, such as the resonating valence bond theory \cite{anderson_resonating_1987} (a failed superconductor), and recent theories proposing topological order for the pseudogap phase \cite{scheurer_topological_2018} by means of failed antiferromagnetism.
In the present work we explore the surprising implications of Eq.~\ref{eq:Sigma} for the topology of strongly correlated electronic systems.
The link between Green's function zeros and interacting topology has been pioneered by Gurarie \cite{gurarie_single-particle_2011} and Volovik \cite{unruh_quantum_2007,volovik_topology_2012} and their role has been considered for the topological classification of various systems \cite{manmana_topological_2012,wang_simplified_2012,slager_impurity-bound_2015,muechler_topological_2020}.
A simple connection between the topological properties of the zeros and the microscopic non-interacting Hamiltonian is however lacking.

\begin{figure}
    \centering
    \includegraphics[width=0.9\linewidth]{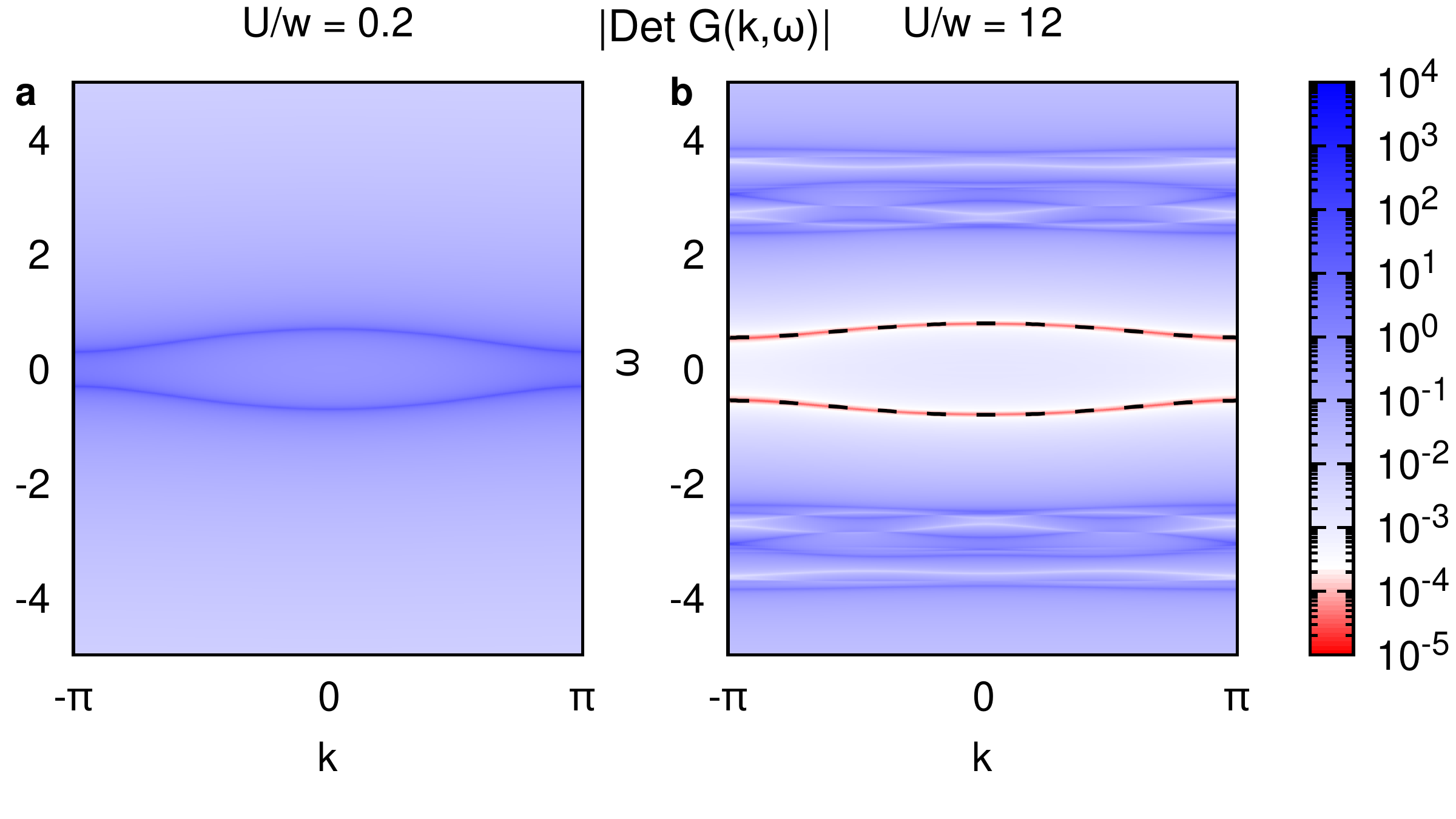}
   	\caption{\textbf{Poles and zeros in an infinite chain.} Spectral representation of the determinant of $G^\mathrm{R}(k,\omega)$ on the real axis highlighting poles (dark blue) and zeros (dark red). In {\bf a}, this is shown for a weakly interacting SSH+$U$ with hopping parameters $v=0.2$ and $w=0.5$ , i.e. on the topologically non-trivial side. 
   	   The numerical solution is obtained with cluster dynamical mean-field theory (CDMFT), whose details can be found in Methods. In the Supplementary Information we show also Quantum Monte Carlo (QMC)  solutions of longer SSH+$U$ chains. 
   	   \textbf{b}, In the strongly interacting limit, in addition to the Hubbard bands at $\pm U/2$, dispersive zeros (in red) are visible inside the spectral gap. The dashed black lines show a fit with the  analytical formula of Eq.~\ref{eq:Sigma} to the position of the zeros of the CDMFT Green's function.}
    	\label{fig:SSH}
\end{figure}

In Fig.~\ref{fig:sketch} we illustrate two prototypical cases predicted by Eq.~\ref{eq:Sigma}: panels (a)-(b) show a symmetry-protected semimetal with bulk bands that meet at some momentum. If the bandwidth is finite, such system in three dimensions turns into a MI at a critical interaction strength \cite{wagner_resistivity_2021}.
As long as the interaction does not break any symmetry that protects the cones of the non-interacting dispersion, Eq.~\ref{eq:Sigma}  dictates that the zeros of $G$ will display a crossing: Analogously to the non-interacting case, a gap in the zeros can only be opened when two ``zero-nodes'' meet in momentum space, which may happen due to the renormalization of the parameters.
This finding offers a particularly transparent explanation of the Mott transition in a Dirac or Weyl semimetal \cite{braguta_monte_2016,wagner_resistivity_2021}:
even if a gap between poles of $G$ opens, the protected linear crossing is in fact not lifted: it just occurs between spectroscopically invisible zeros. 
The second implication regards a TI with an inverted bulk gap and boundary Green's function poles, as shown in Fig.~\ref{fig:sketch}c-d, turning into a topological Mott insulator (TMI) at large interaction strengths. The bulk zeros of $G$ responsible for the opening of the Mott gap again obey Eq.~\ref{eq:Sigma}: depending on the renormalization of $\widetilde{H_0}({\bf k})$, the zeros can acquire an inverted gapped dispersion. 
The sketch also illustrates that the predicted gapless zeros are spatially localized at the boundaries. Their dispersion inside the bulk gap follows the renormalized one of the edge modes in the corresponding non-interacting TI with opposite sign.

In Supplementary Note 1E we give a proof that the topological invariant of a Mott insulator is fully determined by the renormalized Hamiltonian $\widetilde{H_0}({\bf k})$ describing the gapped zeros.
Hence, our approach represents a particularly flexible way of diagnosing all those interacting topological phases that can be classified via one-particle quantities.
It is simple to see that Eq.\eqref{eq:Sigma} recovers the atomic limit when the renormalization of the hopping terms in $\widetilde{H_0}$ becomes zero.
In the following we compare the predictions of Eq.~\ref{eq:Sigma} with numerical results, focusing on a simple one-dimensional case. Additional numerical evidence for the applicability of our approach in two and three dimensions is given in Supplementary Note 3 and Supplmenetary Fig.21-24, where we show various results for the cases sketched in Fig.~\ref{fig:sketch}.

\bigskip
\noindent	
{\bf Infinite and finite SSH+U chains} -- 
Firstly we test the validity of Eq.~\ref{eq:Sigma} with the example of a one-dimensional Su-Schrieffer–Heeger (SSH) model \cite{su_soliton_1980} with periodic boundary conditions.The short bond hopping parameter is set to $v=0.2$ and the long bond to $w=0.5$. We supplement the model  by a local Hubbard $U$ in order to induce the Mott phase. 
For an infinite SSH+$U$ chain at small values of $U$, most of the spectral weight $A({\bf k},\omega)$ resembles the non-interacting eigenvalues $|v+\operatorname{e}^{-i k} w|$ (blue bands in Fig.~\ref{fig:SSH}a).
At large interactions instead, a gap of order $U$ is sustained by low-energy divergencies of $\Sigma({\bf k},\omega)$ (red dispersive features in Fig.~\ref{fig:SSH}b). 
This result, which is in agreement with previous literature \cite{manmana_topological_2012,yoshida_characterization_2014}, allows us to compare the momentum dispersion of the zeros of $G$ with those predicted by Eq.~\ref{eq:Sigma}, shown by the black dashed lines in Fig.~\ref{fig:SSH}b. 
The renormalized dispersion of the zeros perfectly describes our numerical result (see Methods), confirming that Eq.~\ref{eq:Sigma} captures the essence of the momentum-dependent self-energy of a MI.

Since our focus is on boundary zeros, we need to validate Eq.~\ref{eq:Sigma} also in the case of an open SSH+$U$ chain. 

\begin{figure*}[th]
		
		\includegraphics[width=0.9\linewidth]{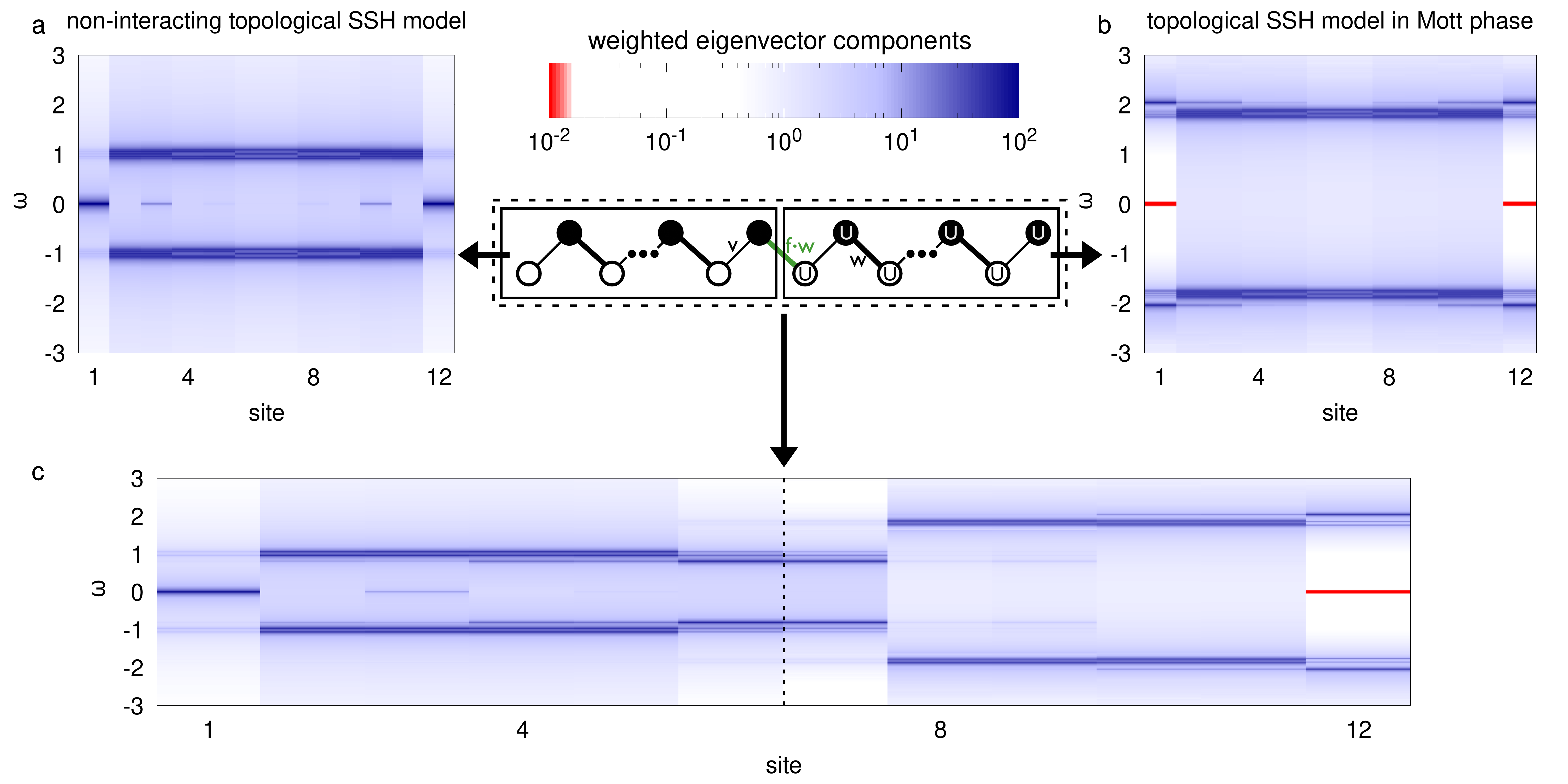}
	
		\caption{\textbf{Zero/pole annihilation in an SSH+$U$ finite chain}.
		\textbf{a}, Exact diagonalization of a 12-site non-interacting SSH chain in the topologically non-trivial phase ($v=0.1$ and $w=1.0$). Zero-dimensional states are visible at the two ends of the chain with some finite  extension inside the chain. We visualize this by showing site-resolved eigenvector components, weighted with the corresponding eigenvalues (see Methods).
		\textbf{b}, Exact diagonalization of a 12-site SSH+$U$ chain in the topological phase with the same $v$ and $w$ but a large value of the on-site interaction ($U/w=4$), i.e. in the Mott phase.
		The ends of the chain now host two zeros inside the Mott gap (of order 4 in units of  $w$).
		The topological invariant of these two chains is the same but the interacting chain has no edge pole, rather only zeros that are clearly  visualized via the weights.  \textbf{c}, Interface (i.e. $f=1$) between the two situations above; the position of the interface is indicated with a dashed, vertical line. In order to align the chemical potential inside the global bulk gap, we subtract the Hartree-shift on the interacting chain.}
		\label{fig:top_vs_topMott}
	\end{figure*}
 
At $U=0$ and for nonzero winding number of the SSH Hamiltonian, $G$ possesses zero-energy poles at the two ends of the chain. This is signaled by the two blue states on the left- and right-most sites of the chain in Fig.~\ref{fig:top_vs_topMott}a where we show the sum of eigenvector components corresponding to the different lattice sites weighted with their eigenvalues (see Methods). 
For large $U$, we solve the finite chain with exact diagonalization (ED) and obtain zeros at the ends of the chain instead, shown in red in Fig.~\ref{fig:top_vs_topMott}b.
Interestingly, the topological nature of the gap of the zeros, analogously to what happens for the poles of $G$, implies the existence of two ``in-gap'' zeros at the boundary of the system. In the Supplementary Information we compare these ED results with those acquired using Eq.~\ref{eq:Sigma}, demonstrating that the analytic formula is also well suited to describe the non-local many-body features of a MI for finite size systems.

 \begin{figure*}[]
		
		\includegraphics[width=0.9\linewidth]{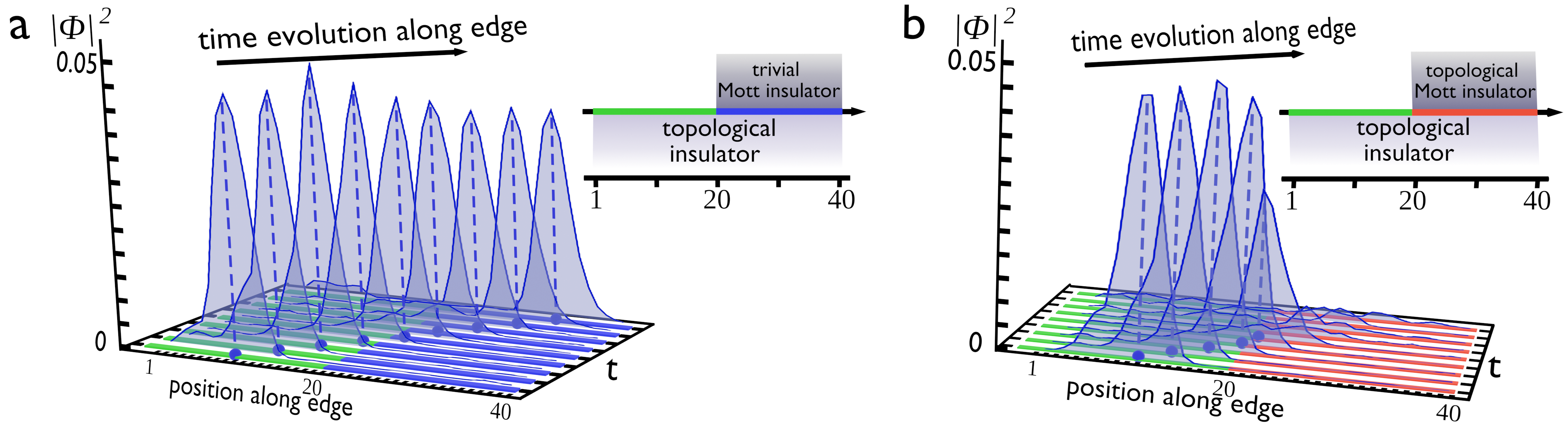}
	
		\caption{\textbf{Evolution of wave packet along the edge.} \textbf{a}, A wave packet is initialized from the leftmost part of the edge of a QSH (green part, see inset) and let evolve towards the interface with a trivial Mott insulator (see inset). Upon entering the blue interface region, no appreciable change of the wave packet is observed, except for some small loss of its relative weight due to reflections and scattering from the walls of the two-dimensional structure.  
		\textbf{b}, The left-moving wave packet evolving along the edge now enters the red region where a topological Mott insulator is placed on the other side. The edge zeros of the TMI hybridize with the edge mode of the QSH and determine the immediate collapse of the wave packet. 
		This sudden loss of weight can be used as a detection protocol for the presence of boundary zeros on the Mott insulating side. 
		In both cases the time evolution is calculated via the Green's function of a 40$\times$40 slab where the interaction is taken into account via the analytic self-energy in Eq.\ref{eq:Sigma}.}
		\label{fig:screenshot_of_the_movie}
\end{figure*}

To address the case in which an edge pole and an edge zero get spatially close to one another we look at the system shown in Fig.~\ref{fig:top_vs_topMott}c. 
The two non-trivial SSH chains ($U$=0 on the left and finite $U$ on the right) are connected by a hopping in the center, that can be switched on and off via a parameter dubbed $f$. 
Fig.~\ref{fig:top_vs_topMott}c shows the fate of pole and zero at the two ends which meet at the center of the new chain. 
They hybridize and, as a result, zeros are no longer located at $\omega=0$. The same happens for the pole localized on the rightmost site of the non-interacting part of the chain.
The solution obtained in Fig.~\ref{fig:top_vs_topMott} is fully compatible with symmetry and topological requirements of having two gapless modes at the two ends of the chain, due to the interchangeable role of poles and zeros.
From a spectroscopic point of view, instead, it is highly unexpected as the new chain, seen as a whole, has in fact only one ``detectable'' low-energy edge state, at the left end. Its partner at the right end, is the dual zero which of course would not be visible in a tunneling experiment. The hybridization does not disappear if pole and zero are at different energies and, in general, if the chiral symmetry is broken. This reflects the fact that both sides of the chain are in the same topological phase and hence there cannot be a protected gapless state localized at the interface. See Supplementary Note 3C for further details.

\bigskip
\noindent
{\bf Two-dimensional topological Mott insulator} --
In two dimensions we find that Eq.~\ref{eq:Sigma} continues to give an excellent description of the dispersion of the zeros of $G$ in MIs. A detailed analysis including a comparison with numerical quantum cluster methods for different real-space geometries, can be found in the Supplementary Information.
Analogously to the one-dimensional case, we define a TMI through the bulk-boundary correspondence in an extended sense: if the zeros of $G$ have a topologically inverted gap in the bulk, then gapless zeros have to appear at the edge. 
In contrast, conventional MIs  do not display robust gapless zeros at their boundaries.

Compared to zero-dimensional poles/zeros of the SSH+$U$, the gapless edge zeros -- living inside the gap of the bulk zeros of a TMI -- acquire a dispersion w.r.t. the momentum parallel to the edge. This implies that the pole/zero annihilation gets even more intriguing than that shown in Fig.~\ref{fig:top_vs_topMott}c. 
In the following we therefore analyze a two-dimensional heterostructure between a conventional quantum spin Hall (QSH) system and a TMI, as illustrated in Fig.~\ref{fig:screenshot_of_the_movie}b. The two parts have a segment of the edge in common and we ask what the consequences are on the helical modes when they travel in this region. 
We also consider two additional cases: a ``benchmark'', in which the two sides are completely disconnected ($f$=0) and one where the TMI is replaced by a trivial Mott (Fig.~\ref{fig:screenshot_of_the_movie}a).
We initialize a wave packet at the very left of the QSH side and let it evolve in time towards the interface with the TMI.
The time evolution is governed by the full Green's function of this hybrid 2D interacting system, where the interaction is included using the analytic formula of Eq.\ref{eq:Sigma}. 
As shown in Fig.~\ref{fig:screenshot_of_the_movie}b as well as in the movie in the Supplementary Information, the wave packet is well defined only up to the  interface with the TMI (marked in red in the inset to Fig.~\ref{fig:screenshot_of_the_movie}b).
As soon as it enters this region, the edge state becomes immediately incoherent and loses spectral weight.
In the other two cases ($f$=0 and trivial MI) the propagation proceeds undisturbed, as standard topological arguments predict (Fig.~\ref{fig:screenshot_of_the_movie}a and movie in the Supplementary Information \cite{noauthor_see_2023}).
The annihilation depends on the relative slope of the edge state and edge zero. If their slope is very similar (as in Fig. \ref{fig:screenshot_of_the_movie}) the reduction of the wave packet weight in the coupled region is maximal. In case of different slopes, the effect remains albeit being quantitatively less pronounced (see Supplementary Note 3G and 3H). The existence and robustness of the gapless edge zeros in the TMI and the observed annihilation with poles on the conventional QSH side is supported by our analytic calculation (Supplementary Note 1E \cite{noauthor_supplementary_2023}), indicating  that deep in the Mott phase, the topological invariant of the interacting phase can be simply calculated via the momentum dependence of $\widetilde{H_0}({\bf k})$.

We have thus described an experimental probe sensitive to the otherwise invisible zeros. 
Based on the dynamics of the wave packet and its clearcut distinct coherence, the propagation represents a  way to detect the presence of the boundary zeros on the TMI side. We have checked that when the wave packet starts to lose weight, there is no component of the QSH edge state that tries to circumvent the TMI part. This is strikingly different from what would instead happen if we were to replace a portion of the QSH with a trivial system. In that case the edge state of the QSH would go around the trivial region. Here instead, the QSH loses the helical state even though  its  bulk topological properties are the same as before. The next question, beyond the scope of the present work, is to understand the consequences of the observed loss of weight for quantum edge transport.

Operatively, one can first quantify the renormalization of parameters of a single-particle tight-binding model for a given material,
as outlined in the Supplementary Information. Then, Eq.~\ref{eq:Sigma} can be used to predict the nature of the corresponding MI. 
Since, in absence of spontaneous symmetry breaking, no higher-order corrections beyond Eq.~\ref{eq:Sigma} can change the symmetry class of the problem \cite{chiu_classification_2016,crippa_fourth-order_2021},  the renormalization of the parameters in $\widetilde{H_0}({\bf k})$ can only influence the position of the topological phase transition.  In other words, the non-trivial region on the Mott side can either be smaller or larger than that of $H_0({\bf k})$, i.e. one can generate a TMI from a trivial band structure or get vice-versa a trivial Mott from a QSH, depending on the specific model and type of interaction. 
Should a Mott material be non-trivial according to Eq.~\ref{eq:Sigma}, one can then exploit the annihilation phenomenon discussed in the last part of this work to unambiguously tell whether or not zeros exist at its boundary.

We have shown that physically detectable (via edge state annihilation)  topological information is carried by the Green's function zeros. It is interesting to connect this information to properties of the full many-body spectrum \cite{fabrizio_spin-liquid_2022}. Investigation of this interesting issue is ongoing but here we show in Sec. 3L of the Supplementary Information, a straightforward connection between the zeros and the spin gap in the case of the simple SSH+U model, where with periodic boundary conditions we show that gapless spin excitations appear exactly when the gap of the zeros closes at the topological phase transition.
A more thorough study of the relation between zeros and spinons beyond the simple case of SSH-model will be part of future work.

These results open therefore interesting perspectives in connection to two-particle indicators of topology \cite{soldini_interacting_2022,herzog-arbeitman_interacting_2022}, Fermi-liquid approaches to the quasiparticles at the Luttinger surface \cite{fabrizio_emergent_2022,fabrizio_spin-liquid_2022} as well as the theory of topological order and in the characterization of protected bulk and surface features in the realm of non-Hermitian physics with strong correlation.

\bigskip
\noindent
{\it Note added in proof}. During the completion of this work, we have become aware of results on the symmetry constraints for the Green's function zeros with the Hatsugai-Kohmoto interaction (see Ref.~\cite{setty_symmetry_2023,setty_electronic_2023}).

\bigskip
\textbf{\large Methods}\\
\textbf{Cluster-DMFT.}
The numerical results for the bulk systems presented in this work are obtained within the framework of cluster-DMFT, an extension of Dynamical Mean-Field Theory capable of grasping nonlocal correlations. The Exact Diagonalization results have been obtained using a cluster extension of the EDIPack code \cite{amaricci_edipack_2022}, where the SSH model is mapped to a finite ``cluster impurity", consisting of two or three interacting dimers, coupled to a finite bath. This is structured as a number of non-interacting clusters replicating the one-particle hopping matrix of the impurity, each site of which is coupled to the corresponding impurity one. The intra-replica hopping amplitudes and bath-impurity couplings are adjusted self-consistently. Two such replicas have been used for the 2-dimers case, and 1 replica for the 3-dimer case.
The BHZ model is solved through an asymmetric cluster impurity consisting of two sites along the $x$ direction, coupled to two bath replicas. Benchmark tests with a 3x1-sites cluster plus 1 bath replica and a 2$\times$2 cluster plus 1 replica have confirmed analogous results, the latter restoring the symmetry of the model though at the cost of a dramatically increased computational time.

\textbf{Single-shot calculation for finite chains.}
The finite size SSH effects have been obtained through a single-shot exact diagonalization of an impurity cluster of 6 dimers decoupled from any bath. For the quantum Monte Carlo calculations a single-shot solution of the impurity problem, again decoupled from any bath, has been acquired via the use of a continuous time quantum Monte Carlo solver based on the interaction expansion (CT-INT)\cite{parcollet_triqs_2015}. For every QMC calculation a statistic of 50 million Monte Carlo cycles is used. 

\textbf{Determination of the finite chain zeros.}
In order to spatially resolve the zeros of real-space Green's function (cfr. Fig. \ref{fig:top_vs_topMott}) we look at weighted eigenvector components, i.e. the quantity
\begin{equation}
    w_i = \sum_j \big\lvert \psi_i^{(j)} E_j\Big\rvert
\end{equation}
where $\psi_i^{(j)}$ is the i-th element of the j-th eigenvector and $E_j$ is the j-th eigenvalue of the Green's function and $i$ and $j$ correspond to lattice sites. This weight is a good indicator of isolated zeros, which are of interest to the topological characterization. On the contrary, it correctly avoids showing the presence of zeros when these are masked by weight coming from other nonzero eigenvalues.
For this reason, the zeros in Fig.~\ref{fig:top_vs_topMott} seem to be entirely localized at the end of the chain without smearing into the inner part of the chain and bulk zeros are also not highlighted in this site-resolved representation.\\
\textbf{Slab wavepacket evolution.}
For the 2D slab calculations the interaction is taken into account using the analytic formula for the self-energy (Eq. \ref{eq:Sigma}). The time evolution of a wave packet at location $r$ and time $t$ is given by
\begin{equation}
	\psi(r,t)=\int d r' G(r,r',t) f(r_0-r')
\end{equation}
where $f(r_0-r')$ is a gaussian centered about $r_0$.

\textbf{Data availability}\\
The data generated in this study have been deposited in the NOMAD database\href{https://doi.org/10.17172/NOMAD/2023.11.06-1}{https://doi.org/10.17172/NOMAD/2023.11.06-1}.

\textbf{Code availabilty}\\
The code used for the CDMFT calculations is available from \href{https://github.com/lcrippa/CDMFT-MOTT-ZEROS}{https://github.com/lcrippa/CDMFT-MOTT-ZEROS}.

\bigskip

%\bibliography{references.bib}
%apsrev4-2.bst 2019-01-14 (MD) hand-edited version of apsrev4-1.bst
%Control: key (0)
%Control: author (8) initials jnrlst
%Control: editor formatted (1) identically to author
%Control: production of article title (0) allowed
%Control: page (0) single
%Control: year (1) truncated
%Control: production of eprint (0) enabled
%

\begin{acknowledgements}
\noindent
\textbf{ Acknowledgments} \\
We thank Jan Carl Budich, Sergio Ciuchi, Michele Fabrizio, Alessandro Toschi and Bj\"orn Trauzettel for useful comments and Lukas M\"uchler for discussions in the early stage of this project.
N.W. is supported by the SFB 1170 Tocotronics,
funded by the Deutsche Forschungsgemeinschaft (DFG, German Research Foundation) – Project-ID 258499086.
L.C. acknowledges financial support from the DFG through the W\"urzburg-Dresden Cluster of Excellence on Complexity and Topology in Quantum Matter–{\it ct.qmat} (EXC 2147, project-id 390858490). 
G.S. acknowledges support from the DFG through FOR 5249-449872909 (Project P05).
We gratefully acknowledge the Gauss Center for Supercomputing e.V. (www.gauss-center.eu) for funding this project by providing computing time on the GCS Supercomputer SuperMUC at Leibniz Supercomputing Center (www.lrz.de). Part of the numerical calculations were carried out using the Julia programming language \cite{bezanson_julia_2017}. We thank the computing service facility of the MPI-FKF for their support. We gratefully acknowledge use of the computational resources provided by the Max Planck Computing and Data Facility. 
We thank José M. Pizarro and the FAIRmat consortium for the support given on the management of our data.  A.A. acknowledges funding from the National Recovery and
Resilience Plan (NRRP) MUR Project PE 0023, CUP B53C22004180005-NQSTI.
J.C. acknowledges support from the Air Force Office of Scientific Research under Grant No. FA9550-20-1-0260.
A.J.M. was supported in part by Programmable Quantum Materials, an Energy Frontier Research Center funded by the U.S. Department of Energy (DOE), Office of Science, Basic Energy Sciences (BES), under Award No. DE-SC0019443. The research leading to these results has received funding from the European Union’s Horizon 2020 research and innovation programme under the Marie Sk{\l}odowska-Curie Grant Agreement No. 897276. The Flatiron Institute is a division of the Simons Foundation. 
\end{acknowledgements}
\textbf{Author contribution}\\
The original idea by G.S. was further discussed with A.G., J.C. and A.M. (first 
in a taxi ride to Tokyo Haneda Airport and then during a visit at CCQ, Flatiron 
Institute). N.W. and L.C. have led the project under the supervision of G.S and 
A.M. The numerical and analytical calculations were performed by N.W., L.C., A.A., 
M.K. and E.K.  All previously mentioned authors as well as P.H, T.S, and D. Di S.  exchanged views on the results and contributed to the 
writing of the manuscript.\\
\textbf{Competing Interest Statement}
The authors declare no competing interests.

\end{document}